\documentclass[aps,prl,twocolumn,showpacs,superscriptaddress,floatfix]{revtex4-1}
\usepackage{graphics}
\usepackage{graphicx}
\usepackage{amsmath}
\usepackage{amsfonts}
\usepackage{amssymb}
\usepackage{pstricks}
\begin{document}

\title{Cross-stream transport of asymmetric particles driven by oscillating shear}

\author{Matthias Laumann}
\affiliation{Theoretische Physik I, Universit\"at Bayreuth, 95440 Bayreuth, Germany}
\author{Paul Bauknecht}
\affiliation{Biofluid Simulation and Modeling, Universit\"at Bayreuth, 95440 Bayreuth, Germany}
\author{Stephan Gekle}
\affiliation{Biofluid Simulation and Modeling, Universit\"at Bayreuth, 95440 Bayreuth, Germany}
\author{Diego Kienle}
\email{corresponding author: diego.kienle@uni-bayreuth.de}
\affiliation{Theoretische Physik I, Universit\"at Bayreuth, 95440 Bayreuth, Germany}
\author{Walter Zimmermann}
\affiliation{Theoretische Physik I, Universit\"at Bayreuth, 95440 Bayreuth, Germany}

\date{\today}

\begin{abstract}
We study the dynamics of asymmetric, deformable particles in oscillatory, linear
shear flow.
By simulating the motion of  a dumbbell, a ring polymer, and a capsule we show
that cross-stream migration occurs for asymmetric elastic particles even in linear
shear flow if the shear rate varies in time. The migration is generic as it does
not depend on the particle dimension.
Importantly, the migration velocity and migration direction are robust to variations
of the initial particle orientation, making our proposed scheme suitable for sorting
particles with asymmetric material properties.
\end{abstract}

\maketitle

{\it Introduction.}
During the recent years, microfluidics has evolved to a cross-disciplinary field, ranging
from basic physics to a plethora of biological and technical
applications~\cite{Squires:RMP2005,Whitesides:Nature2006,Popel:ARFM2005,
Graham:ARFM2011,Dahl:ARCBE2015,Sackmann:Nature2014,Amini:LC2014},
including the control of small amounts of fluids, chemical
synthesis~\cite{Jaehnisch:ACIE2004,Elvira:NatureChe2013}, biological
analysis~\cite{Yi:ACA2006,Chen:LC2012}, and the study of the deformation dynamics of droplets,
vesicles, capsules, or blood
cells~\cite{Shafer:BPC1974.P1P2,Aubert:JCP1980.P1P2,Nitsche:AlChE1996,Ghigliotti:PRL2011,
Sekhon:JPS1982,Brunn:IJMF1983,Jhon:JPS1985,Brunn:RA1984,Agarwal:CES1994,
Cantat:PRL1999,Seifert:PRL1999,Ma:PF2005,
Haber:JFM1971,Leal:ARFM1980,Mandal:PRE2015,
Kaoui:PRE2008,Danker:PRL2009,Doddi:IJMF2008,Sibillo:PRL2006,Abkarian:SM2008,
Dupire:PRL2010,Baroud:LC2010,Deschamps:PNAS2009,
Kessler:EPJE2009,Noguchi:PRE2010,Zhao:PF2011,Matsunaga:JFM2015}.
An important transport mechanism in microfluidic flows is the cross-stream
migration (CSM), where particles move across streamlines and can be sorted
due to their particle-specific properties~\cite{Sajeesh:MFNF2014,Geislinger:ACIS2014}.

The CSM effect has been first reported in 1961 by Segre and Silberberg for
rigid particles at finite Reynolds number in pipes with diameters of several
millimeters~\cite{Segre:Nature1961}.
When channels approach the micrometer scale, the Reynolds number vanishes
(Stokes regime) and fluid inertia does not matter;
likewise, for $\mu\mbox{m}$-sized particles thermal effects can be discarded.
In the Stokes regime, CSM arises in
curvilinear~\cite{Shafer:BPC1974.P1P2,Aubert:JCP1980.P1P2,Nitsche:AlChE1996,Ghigliotti:PRL2011}
and
rectilinear flow~\cite{Sekhon:JPS1982,Brunn:IJMF1983,Brunn:RA1984,Jhon:JPS1985,Agarwal:CES1994},
if the particle is elastic and, in case of rectilinear flow, the flow's fore-aft
symmetry is broken, requiring intra-particle hydrodynamic
interaction~\cite{Sekhon:JPS1982,Brunn:IJMF1983,Brunn:RA1984,Jhon:JPS1985}.
Such symmetry breaking occurs near boundaries via wall-induced lift
forces~\cite{Jhon:JPS1985,Cantat:PRL1999,Seifert:PRL1999,Ma:PF2005}
or by space-dependent shear rates, so that
dumbbells~\cite{Sekhon:JPS1982,Brunn:IJMF1983,Brunn:RA1984},
droplets~\cite{Haber:JFM1971,Leal:ARFM1980,Mandal:PRE2015},
vesicles and capsules~\cite{Kaoui:PRE2008,Danker:PRL2009,Doddi:IJMF2008}
exhibit CSM even in unbounded flow.
These parity breaking mechanisms may be accompanied by other effects due
to viscosity contrast~\cite{Haber:JFM1971,Farutin:PRL2012} or particle
chirality~\cite{Watari:PRL2009}, which further impact the CSM.

Here we show that a controlled cross-stream migration is possible even in unbounded
{\it linear} shear flow, provided that
(1) the particle holds an intrinsic asymmetry (parity breaking), and
(2) the shear rate varies in time, causing time-dependent particle deformations.
Importantly, the cross-stream migration occurs irrespective of the dimensionality
of the particle, accentuating its {\it generic} nature, as we show by studying
particles extending in one (1D), two (2D), and three (3D) dimensions.
We demonstrate that the CSM depends on external flow parameters such as switching
period, which can be controlled conveniently to achieve an optimized migration.

\begin{figure}[!htbp]
\centering
\includegraphics[width=0.95\columnwidth]{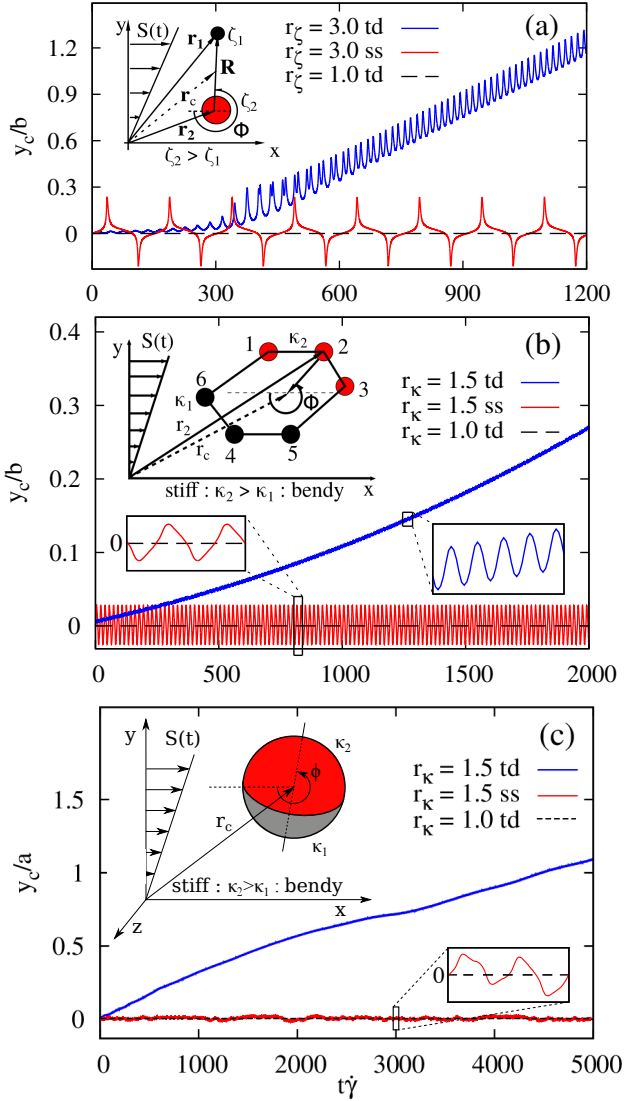}
\vspace{-0.4cm}
\caption{(Color online) Lateral position $y_c(t)$ vs. scaled time
$t\dot\gamma$ for the asymmetric (a) 1D dumbbell, (b) 2D ring, and
(c) 3D capsule, sketches of which are shown in the inset.
Irrespective of the model details, all particle types perform a
{\it net} cross-stream migration in linear shear (blue solid), if
the shear rate is time-dependent (td).
At steady shear (ss), $y_c(t)$ oscillates around a constant mean
(red solid), so that the net migration vanishes~\cite{Brunn:IJMF1983,Nitsche:AlChE1996}.
For symmetric particles ($r_{\zeta,\kappa} = 1.0$), the migration
is zero.
The initial orientation is $\phi_0 = 2.0\pi$.}
\label{Fig_Migration_Generic}
\end{figure}

{\it Approach.}
To reveal the generic behavior of the CSM in oscillatory shear flow, we use three
kinds of particles, which share the common features that they are deformable,
asymmetric, and their constituent parts interact hydrodynamically.
The first two particle types are a dumbbell (1D) and a ring-polymer (2D), modeled by
a sequence of bead-spring units and connected by linear springs with a bond length $b$
and force constant $k$.
The dumbbell asymmetry is modeled by assigning different friction coefficients
$\zeta_1$ and $\zeta_2$ to unequal sized beads $1$ and $2$ with
$r_{\zeta} = \zeta_2/\zeta_1 = 3$ [Fig.~\ref{Fig_Migration_Generic} (a): inset].
The asymmetry of the $N$-bead ring-polymer is realized by a space-dependent bending
stiffness $\kappa \left( \left\{ {\bf r} \right\} \right)$ along the ring contour
$\left\{ {\bf r} \right\}$.
The third particle is an elastic capsule (3D), the asymmetry of which is implemented
likewise by a spatially varying bending stiffness
$\kappa \left( \left\{ {\bf r} \right\} \right)$ along the capsule surface
$\left\{ {\bf r} \right\}$.
For the purpose of this study, we split the contour/surface of the ring/capsule in
equal parts (Janus-particle); the stiff and bendy portion in either case has a
bending stiffness of $\kappa_2$ and $\kappa_1$ with a ratio 
$r_{\kappa} = \kappa_2/\kappa_1 = 1.5$ [Fig.~\ref{Fig_Migration_Generic} (b) and (c): inset].

The migration dynamics of all three particle kinds is obtained from their
non-Brownian trajectories.
The trajectories for the dumbbell and ring-polymer are determined by solving
the Stokesian dynamics for each bead,
including the hydrodynamic backflow (HB) due to the fluid-particle interaction
(HI) described by the Oseen tensor~\cite{DoiEd:1986}.
The capsule path is calculated using the Immersed Boundary Method in conjunction
with the Lattice Boltzmann method for the flow~\cite{Krueger:CMA2011}, employing
an adapted version of the ESPResSo package~\cite{LBM:Espresso}.
Throughout we assume a time-dependent (td), linear shear flow
${\bf u}\left( x,y\right) = S\left( t \right) y {\bf e}_x$
along the ${\bf e}_x$-axis;
the shear rate $S(t)$ has a period $T$ with $S(t) = +\dot \gamma$ during the
first half-period $T_1$ and $S(t) = -\dot\gamma$ during the second half-period
$T_2$ with $T_1=T_2=T/2$.
The initial orientation of all three particles is $\phi_0 = 2.0\pi$ with the
small $\zeta_1$-bead, respectively, the stiff $\kappa_2$-contour/surface being
located to the left.
The Supporting Information~\cite{RefSI} provides further details.

{\it Generic Behavior.}
Figure~\ref{Fig_Migration_Generic} shows the center-of-mass position $y_c (t)$,
scaled with respect to the dumbbell/ring bond length $b$ or the capsule radius
$a$, as a function of the scaled time $t\dot\gamma$ with fixed $\dot\gamma$ for
all three particles.
For symmetric particles ($r_{\zeta,\kappa} = 1.0$), the cross-stream migration
is zero at any time (dashed line)~\cite{Brunn:IJMF1983,Nitsche:AlChE1996} as
parity breaking does not occur irrespective of whether the shear flow is at
steady-state (ss) or time-dependent (td).
For asymmetric particles ($r_{\zeta,\kappa} > 1.0$) in steady shear flow parity
breaking exists, but which results only in an alternating CSM as reflected in the
oscillatory behavior of $y_c(t)$; the {\it net} migration over one shear cycle,
however, is still zero (red solid line).
This intermittent migration of asymmetric particles can be exploited to attain
a {\it net} cross-stream migration, if the shear rate $S(t)$ is made time-dependent
by switching $S(t)$ at a frequency $1/T$, as shown in Fig.~\ref{Fig_Migration_Generic}
(a)-(c) by the blue solid line.
The fact that all three particles display cross-stream behavior irrespective of their
dimensionality and model details is an indication of a {\it generic} property~\cite{RefSI},
which can be attributed to the different {\it mean} shapes the particle acquires during
each half-period, as discussed next.

\begin{figure}[htbp]
\centering
\includegraphics[width=0.9\columnwidth]{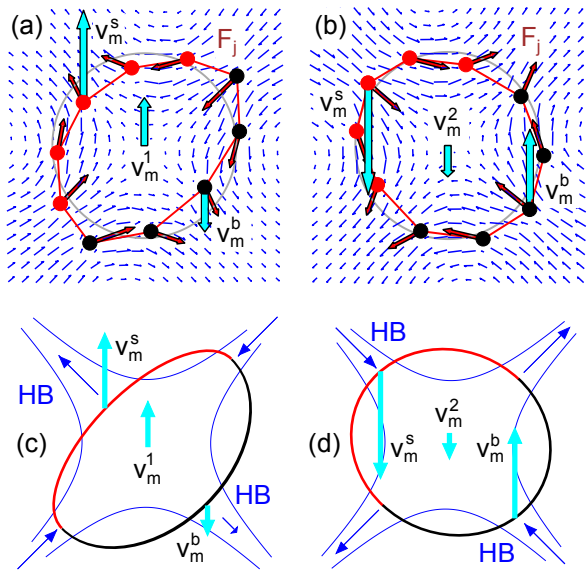}
\vspace{-0.3cm}
\caption{(Color online)
Explanation of the basic migration mechanism by way of the ring.
Shown are simulation data for the {\it mean} ring shape, averaged over the
(a) first $T_1$ and (b) second $T_2$ shear cycle, and the potential forces
${\bf F}_j$ along the $N$-bead contour.
While the backflow ${\bf v} ({\bf r})$ (HB) induced within each half-cycle
$T_i$ is approximately elongational (blue), its parity is broken due to the
particle asymmetry and leads to different CSM-driving drags $v_{m}^{s}$ and
$v_{m}^{b}$ at the stiff and bendy side.
Fig.~(c) and (d) show how the parity asymmetry of the HB is enhanced
($|v_{m}^{s}| > |v_{m}^{b}|$) during the $T_1$-cycle, while during the
$T_2$-cyle the HB-parity is partially recovered ($|v_{m}^{s}| \succsim |v_{m}^{b}|$),
leading to a {\it net} positive migration $v_m = v_{m}^{1} + v_{m}^{2} > 0$.}
\label{Fig_Migration_Explain}
\end{figure}

{\it Migration Mechanism.}
To understand the CSM mechanism we use the $N$-bead ring as representative particle
and consider the case, where the ring has approached the steady-state, i.e., for an
initial orientation $\phi_0$ the ring adopts (after a transient) one stable {\it mean}
orientation $\langle \phi \rangle_T$ over one shear-cyle $T$.
Specifically, for $\phi_0 = 2.0~\pi$ the mean orientation
$\langle \phi \rangle_T \approx 1.75~\pi$ with the stiff contour located in the upper
(left) half-space and referred to below.
We analyze the migration in terms of the {\it mean} steady-state CSM velocity $v_{m}^{i}$
(along the $y$-axis) for each half-period $T_i$ by averaging the velocity $v_m (t)$ over
$T_i$~\cite{RefSI}
\begin{equation}\label{Eq_VelCSM}
v_{m}^{i} = \langle {\bf e}_y \cdot \dot{\bf r}_c (\infty) \rangle_{T_i} =
\frac{1}{N} \sum_{i=1}^{N} \sum_{j \neq i}
\langle {\bf e}_y \cdot {\bf H}_{i,j} \cdot {\bf F}_j \rangle_{T_i}~.
\end{equation}
Eq.~(\ref{Eq_VelCSM}) implies that for rectilinear flows with ${\bf e}_y \cdot {\bf u} = 0$,
the cross-stream transport is entirely driven by the particle drag due to the hydrodynamic
backflow, induced by the potential forces ${\bf F}_j$.
Further, the magnitude and direction of each HB (and hence of $v_{m}^{i}$) depend on the
particle shape via the force profile ${\bf F} \left( \left\{ {\bf r}_j \right\} \right)$
and the dyadic mobility matrix ${\bf H}_{i,j}$.
An expression similar to Eq.~(\ref{Eq_VelCSM}) can be used to determine the {\it mean}
HI-induced flow field ${\bf v}({\bf r})$ for each half-period~\cite{RefSI}.
The respective 2D backflow ${\bf v}({\bf r})$, shown in Fig.~\ref{Fig_Migration_Explain}
(a) and (b) when $\langle \phi \rangle_T \approx 1.75~\pi$, corresponds to an {\it elongational}
flow, which flow lines are reversed (sign change) as a result of the altering ring deformation
during the $S(t)$-switching ($+\dot\gamma \rightarrow -\dot\gamma$), and displayed more clearly
in Fig.~\ref{Fig_Migration_Explain} (c) and (d).
The ring asymmetry causes generally a break of the parity symmetry (PS) of the elongational
HB, but the extent of the PS-violation depends on the strength of the {\it mean} deformation
during each half-cycle $T_i$ [Fig.~\ref{Fig_Migration_Explain} (c) and (d)].
 
Comparing the {\it mean} deformation for each half-cyle, one observes that the ring asymmetry
is {\it enhanced} during the first $T_1$ shear-cycle, causing an increased parity break
of the elongational HB [Fig.~\ref{Fig_Migration_Explain} (c)].
But this implies that the {\it difference} between the opposing partial HB-drags at the stiff
and bendy side, $v_{m}^{s}$ and $v_{m}^{b}$ (see \cite{RefSI}: Eq.~(9)), becomes larger with
$|v_{m}^{s}| > |v_{m}^{b}|$ since $\kappa_2 > \kappa_1$.
Hence, the {\it mean} migration step $v_{m}^{1} = v_{m}^{s} + v_{m}^{b}$ during the $T_1$-cycle
is large and {\it positive}.
In turn, during the $T_2$ shear-cyle the situation is reversed as the ring shape is roughly
circular, i.e., the ring asymmetry is {\it reduced} with the result that the parity of the
elongational backflow is partly recovered [Fig.~\ref{Fig_Migration_Explain} (b) and (d)],
and the opposing partial HBs almost cancel.
The reason for the residual backflow is because the HB-drag at the stiff contour part is slightly
larger than at the bendy side ($|v_{m}^{s}| \succsim |v_{m}^{b}|$), as a result of the larger stiffness.
During the $T_2$-cycle the {\it mean} migration step $v_{m}^{2} = v_{m}^{s} + v_{m}^{b}$ is thus
small and {\it negative}.
Over the course of one shear-cyle $T$, the {\it net} migration $v_m = v_{m}^{1} + v_{m}^{2}$
is therefore {\it positive}, as displayed by all three kinds of particles
[Fig.~\ref{Fig_Migration_Generic} (a)-(c)].

\begin{figure}[!htbp]
\centering
\includegraphics[width=1.0\columnwidth]{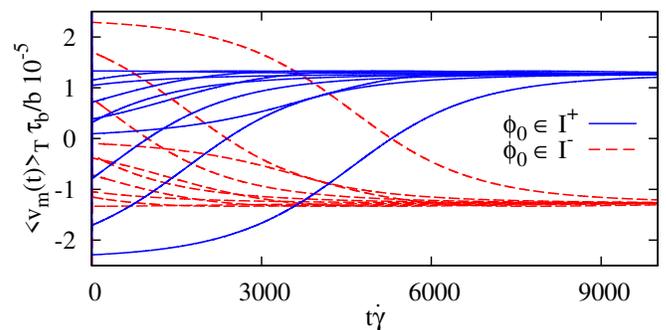}
\vspace{-0.8cm}
\caption{(Color online) Migration velocity
$\langle v_m (t) \rangle_{T} \tau_b / b$ of the ring, averaged over one
period $T$, versus time $t\dot\gamma$ for different initial angles $\phi_0$
taken out of the interval $I^{+} = \left[1.1;2.0\right]\pi$ or
$I^{-} = \left[0.1;1.0\right]\pi$.
The $\langle v_m (t) \rangle_{T}$-transient depends on $\phi_0$, while
the steady-state value $v_m = \langle v_m (\infty) \rangle_{T}$ is
$\phi_0$-independent.
Parameters: $\dot\gamma = 0.1$, $T = 20$, $r_{\kappa} = 1.5$.}
\label{Fig_RingPoly_Vm_t}
\end{figure}

{\it Orientation Robustness.}
We now demonstrate that the CSM effect is quite robust against a dispersal
of initial orientations by varying the angle $\phi_0$, while keeping the
orientation axis within the $y$-$x$ shear plane.
Figure~\ref{Fig_RingPoly_Vm_t} shows the migration velocity of the ring
$\langle v_m (t) \rangle_{T} \tau_b / b$, averaged over one cycle $T$,
versus time $t\dot\gamma$ ($\dot\gamma=0.1$ fixed,
$\tau_b = \eta b^3 / \kappa_1$) for various orientations $\phi_0$ within
the intervals $I^{+} = \left[1.1;2.0\right]\pi$ and $I^{-} = \left[0.1;1.0\right]\pi$.
The interval $I^{+}$ ($I^{-}$) corresponds to ring orientations, where during
migration the stiff/bendy part lies in the mean within the upper/lower (lower/upper)
half-space.
As discussed before, this implies that the final, steady-state migration velocity
$v_m \equiv \langle v_m (\infty) \rangle_{T}$, is positive for $\phi_0 \in I^{+}$
and negative for $\phi_0 \in I^{-}$, as disclosed in Fig.~\ref{Fig_RingPoly_Vm_t}
for $t\dot\gamma > 9.0\cdot 10^3$.
Remarkably, the ring migrates always at the same steady-state speed $v_{m}^{+/-}$
even though the $\phi_0$-orientation varies by almost $\Delta\phi_{0}^{+/-} \approx \pi$,
meaning $v_{m}^{+/-}$ is independent of $\phi_0$.
In turn, the choice of $\phi_0$ determines strongly the short-time dynamics of
$\langle v_m (t) \rangle_{T}$, as shown in Fig.~\ref{Fig_RingPoly_Vm_t} for
$t\dot\gamma < 9.0\cdot 10^3$.
This imbalance of the magnitude and in part the sign of $\langle v_m (t) \rangle_{T}$
is a transient signature and exists as the orientation angle $\phi (t)$ is not
yet in-phase with the shear signal $S(t)$;
the phase synchronization of the angle takes place gradually within the transient
regime over many shear cycles $T$ before a phase-locking is established.
The behavior of $\langle v_m (t) \rangle_{T}$, shown for the ring in
Fig.~\ref{Fig_RingPoly_Vm_t}, is generic and displayed by the other
particles types~\cite{RefSI}.
We note that the migration persists ($v_m \neq 0$) when the tilt angle $\theta_0$
between the particle axis and the $y$-$x$ shear plane is non-zero~\cite{RefSI},
accentuating the robustness of the CSM effect against orientation variations.
For particle orientations with $\theta_0 = \pi/2$ the upper-lower symmetry prohibits
migration ($v_m = 0$) for any $\phi_0$-value.

\begin{figure}[htb]
\centering
\includegraphics[width=1.0\columnwidth]{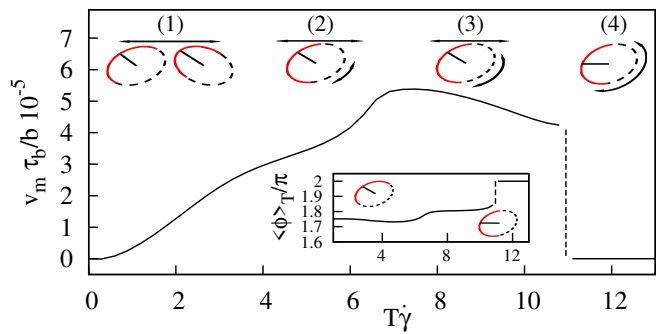}
\vspace{-0.7cm}
\caption{(Color online) Migration velocity $v_m \tau_b / b$ of the ring
versus switching period $T\dot\gamma$.
Four dynamic regimes are identified:
(1) oscillatory shear deformation, indicated by the horizontal arrow, at
small $T$;
(2) weak tank-treading (TT) superposed with (1), marked by the half-circle arrow;
(3) enhanced tank-treading at large $T$;
(4) TT-dominated with zero net migration for $T\dot\gamma > 11$.
Inset: $T$-averaged mean orientation $\langle \phi \rangle_T$ vs. $T\dot\gamma$.
Parameters: $\phi_0 = 2.0~\pi$, $\dot\gamma = 0.1$, $r_{\kappa} = 1.5$.}
\label{Fig_RingPoly_Vm_Tgam}
\end{figure}

{\it Frequency Dependence.}
The migration process is not entirely determined by the material properties of
the particle (e.g. stiffness), but can be controlled also by external parameters
such as the shear rate $\dot\gamma$ or the switching period $T$, the latter being
discussed next.
Figure~\ref{Fig_RingPoly_Vm_Tgam} shows the steady-state migration velocity
$v_m \tau_b / b$ for a fixed $\dot\gamma = 0.1$ versus the period $T\dot\gamma$,
which sets the time scale for the sign change of the shear rate $S(t)$.
When $T\dot\gamma$ is small, the migration speed $v_m$ is rather low (regime (1))
since the quickly alternating shear rate $S(t)$ induces only a small shear deformation
of the ring shape, so that the ring has not sufficient time to reorient and to
fully develop its mean conformation within each half-period $T_1$ and $T_2$, respectively;
at these short times tank-treading is still marginal, as sketched in Fig.~\ref{Fig_RingPoly_Vm_Tgam}.
For larger periods $T$, the ring has now more time within each half-cycle to deform
and fully adopt the migration state, so that $v_m$ monotonously grows first (regime (2)),
approaching a maximum at $T\dot\gamma \approx 7$.
At this stage, a weak partial tank-treading (TT) of the contour is initiated, but the
ring dynamics is still dominated by oscillatory shear deformations, driving the CSM.
Beyond a value of $T\dot\gamma > 7$ the migration gradually decays since tank-treading
becomes increasingly important insofar as a larger fraction of time of each half-cycle
$T_{i}$ is spent on tank-treading.
This implies that a portion of the stiff/bendy contour is now partly shuffled from the
upper/lower half-space to the lower/upper one (regime (3)), i.e., the dynamics of the
entire shear cycle takes now place within two half-spaces (with an unequal amount) and 
each contributes to the CSM with opposite sign.
The {\it net} velocity $v_m$ is still positive, since the {\it mean} orientation of the
ring $\langle\phi\rangle_T \approx 1.75\pi$ [Fig.~\ref{Fig_RingPoly_Vm_Tgam}: inset] with
the stiff/bendy contour part residing on {\it average} within the upper/lower half-space.
Beyond $T\dot\gamma > 11$ the CSM comes to a halt since tank-treading dominates now the
dynamics within each half-period, so that even a larger fraction of the stiff/bendy contour
is re-shuffled between the upper-lower half-space.
Within this TT-dominated regime, the {\it mean} orientation flips from 
$\langle\phi\rangle_T \approx 1.75~\pi$ ($v_m > 0$) to
$\langle\phi\rangle_T \approx 2.0~\pi$ [Fig.~\ref{Fig_RingPoly_Vm_Tgam}: inset],
which corresponds to a {\it symmetric} state where equal amounts of the stiff/bendy contour
lie in the {\it mean} within both half-spaces, so that $v_m$ is zero~\cite{RefSI}.
The abrupt $v_m$-drop is hence inherently connected with the abrupt change of the {\it mean}
orientation $\langle \phi \rangle_T$, which can be understood by inspecting the phase-space
$\langle \dot\phi (t) \rangle_T$-$\langle \phi (t) \rangle_T$ (see SI for details). 
Here we just note that the phase-space features a pattern of discrete, {\it asymptotically stable}
orientations $\langle \phi \rangle_{T}^{FP}$ (fixed points), which the ring can access.
Importantly, the number and value of available $\langle \phi \rangle_{T}^{FP}$ depend
sensitively on the switching period $T$~\cite{RefSI}.
In our case with $\phi_0 = 2.0~\pi$, the only stable orientation the ring can adopt is
$\langle \phi\rangle_T \approx 1.75~\pi$ as long as $T\dot\gamma < 11$ while
$\langle \phi\rangle_T \approx 2.0~\pi$ is unstable.
This stability sequence is reversed when $T\dot\gamma > 11$ and the ring locks in to the now
stable mean orientation $\langle \phi\rangle_T \approx 2.0~\pi$ [Fig.~\ref{Fig_RingPoly_Vm_Tgam}: inset].

{\it Conclusions.}
We have shown that deformable particles, which hold an intrinsic asymmetry
(parity breaking), display cross-stream migration (CSM) in time-perodic,
linear shear flow.
The net migration can be attributed uniquely to the particle asymmetry as
it leads to an asymmetric force distribution within the periodically deformed
particle, inducing asymmetric, non-compensating hydrodynamic backflows (HB).
Since the magnitude and direction of the HBs depend on the actual particle
deformation, which is different within the first and second half-period, the
HBs averaged over one shear cycle $T$ are nonzero, thus leading to a finite
CSM [Fig.~\ref{Fig_Migration_Explain}].
The CSM is {\it generic} inasmuch as it does not depend on the particle dimension
nor on the specific details of its asymmetry [Fig.~\ref{Fig_Migration_Generic} (a)-(c)].
While the migration direction is sensitive to whether the stiff/bendy part of the
particle resides during one shear cycle in the mean within the upper or lower half
space, the CSM speed approaches after a transient phase a constant value and is
independent of the initial particle orientation [Fig.~\ref{Fig_RingPoly_Vm_t}].

Given that even a small asymmetry in the bending modulus (factor $1.5$ or less)
of micron-sized particles can trigger a sizable migration velocity of
$20~\mu\mbox{m/min}$ under realistic flow conditions with a shear rate of
$\dot\gamma = 22~\mbox{s}^{-1}$ and a period of $T=1.75~\mbox{Hz}$~\cite{RefSI},
our proposed scheme facilitates appreciable migration distances in compact
microfluidic setups just by independently tuning the amplitude and frequency of
the shear rate.
Investigating effects due to random material inhomogeneities will be an interesting
subject for future studies.

{\it Acknowledgment.}
PB and SG thank the Volkswagen foundation for support and gratefully acknowledge
the Leibniz Supercomputing Center Munich for the provision of computing time.
ML and WZ acknowledge support by the DFG priority program on Micro- and Nanofluidics.

\end{document}